\documentclass[prd,a4paper,aps,epsfig,showpacs,twocolumn,floats]{revtex4}
\usepackage{epsfig}

\newcommand{\bk}{{\mathbf k}}

\newcommand{\bx}{{\mathbf x}}

\newcommand{\bn}{{\mathbf n}}
\newcommand{\bv}{{\mathbf v}}
\newcommand{\bw}{{\mathbf w}}
\newcommand{\bE}{{\mathbf E}}
\newcommand{\bnabla}{{\mathbf \nabla}}

\newcommand{\cd}{\cdot}
\newcommand{\al}{\alpha}

\newcommand{\de}{\delta}

\newcommand{\ga}{\gamma}

\newcommand{\La}{\Lambda}
\newcommand{\la}{\lambda}
\newcommand{\Om}{\Omega}
\newcommand{\om}{\omega}

\newcommand{\dt}{\ensuremath{\delta \theta}}
\newcommand{\dx}{\ensuremath{\delta x}}

\newcommand{\be}{\begin{equation}}
\newcommand{\ee}{\end{equation}}

\newcommand{\bea}{\begin{eqnarray}}
\newcommand{\eea}{\end{eqnarray}}
\newcommand{\bean}{\begin{eqnarray*}}
\newcommand{\eean}{\end{eqnarray*}}
\newcommand{\dd}{\partial}

\newcommand{\ie}{{\em i.e. }}

\newcommand{\hJ}{\hat{J}}
\newcommand{\Sp}{{S'}}
\newcommand{\bt}{\mbox{\boldmath $\theta$}}
\newcommand{\bxi}{\mbox{\boldmath $\xi$}}
\newcommand{\bal}{\mbox{\boldmath $\alpha$}}
\newcommand{\mA}{\mathcal{A}}
\newcommand{\kp}{\kappa_\Psi}
\newcommand{\kv}{\kappa_\bv}
\newcommand{\cv}{C_\ell^\bv}
\newcommand{\cp}{C_\ell^\Psi}

\newcommand{\bb}{\mbox{\boldmath $\beta$}}

\newcommand{\pa}{\ensuremath{\partial}}
\newcommand{\apt}{{\ensuremath{\mathcal H}}}

\begin{document}

\title{Effect of Peculiar Motion in Weak Lensing}

\author{Camille Bonvin}
\email{camille.bonvin@unige.ch}
\affiliation{
D\'epartement de Physique Th\'eorique, Universit\'e
de Gen\`eve, 24 quai Ernest Ansermet, CH--1211 Gen\`eve 4,
Switzerland.}

\date{\today}

\begin{abstract}

We study the effect of peculiar motion in weak gravitational lensing. We derive a fully relativistic formula
for the cosmic shear and the convergence in a perturbed Friedmann Universe. We find 
a new contribution related to galaxies' peculiar velocities. This contribution does not affect 
cosmic shear in a measurable way, since it is of second order in the velocity.
However,  its effect on the convergence (and consequently on the magnification, which is a measurable quantity)
is important, especially for redshifts $z\leq 1$. As a consequence, peculiar motion modifies also the relation
between the shear and the convergence.

\end{abstract}


\maketitle                                                                   

\section{Introduction}
\label{sec:intro}

Mapping the large-scale structure of the Universe is one of the most important current challenges for cosmology.
Weak gravitational lensing represents a promising tool to achieve this goal, since it is directly sensitive to 
the distribution of matter in the Universe, independent of its nature (baryon, dark matter...).
Gravitational lensing describes indeed the deflection of light rays from distant sources by the gravitational potential
along the line of sight.
It induces consequently a modification of the shape of the sources. This distortion of images contains information
about the evolution of large-scale structure, \ie about the geometry and dynamics of the Universe
(see e.g.~\cite{theory,physrep2008} and references therein). Weak gravitational lensing can be divided in two parts: the shear, that distorts the shape of 
the source; and the convergence, that magnifies or demagnifies it.                                                         
Both of these effects have already been measured.
 
Cosmic shear is detected through the correlations it induces on the ellipticity of galaxies.
It was measured for the first time in 2000, by four independent teams~\cite{first_detection_shear}.
Since then, many other experiments have detected cosmic shear in random patches of the sky~\cite{obs_shear,Fu}.
In the next few years,
weak lensing surveys, like CFHTLS~\cite{CFHTLS}, the Dark Energy Survey~\cite{DES} and Pan-STARRS~\cite{pan} will deliver
accurate measurements ($\lesssim1\,\%$ level) over large parts of the sky. In the further future, even more
challenging experiments like Euclid~\cite{Euclid},
LSST~\cite{LSST} and SNAP~\cite{SNAP} are planned. 

The other component of weak lensing, the convergence, can be detected through the modifications it induces
on the galaxy (or quasar) number density over a given flux threshold~\cite{convergence}. 
The convergence (or more precisely the magnification) has already been robustly detected using quasar-galaxy correlations
(see e.g.~\cite{sdss} and references therein).
Moreover, recently~\cite{pen} has highlighted the possibility to measure accurately the magnification autocorrelations with the
Square Kilometer Array (SKA)~\cite{SKA}. A precision of $10\,\%$ is expected at the beginning and $\lesssim1\,\%$ later on. 
Hence the convergence provides an additional precise observational quantity, useful to constrain cosmology.
In order to make optimal use of this observational information, one needs to understand the
underlying theory of weak lensing accurately. 

In this paper, we present a fully relativistic description of weak gravitational lensing.
More precisely, we calculate the Jacobi map, that relates the surface of a galaxy to its angular image at the observer position, following the formalism presented
in~\cite{seitz}. This map describes the distortion of a light beam by density perturbations along the geodesic between the source and the observer. The shear and the convergence are then extracted from this application. Our derivation differs from the standard one in two points.

First, in the usual derivation the $4\times 4$  Jacobi map is reduced to a $2\times 2$ matrix, called the magnification matrix, by assuming that the source and the image belong to the same two-dimensional subspace, normal to the photon direction and to the observer four-velocity. The convergence is then defined as the trace of the magnification matrix and the shear as the traceless part. However, the source four-velocity differs generally from the observer four-velocity. As a consequence
the source and the image do not belong to the same two-dimensional subspace. In this paper we take into account this difference. We show that the Jacobi map can not be reduced to a $2\times 2$ matrix but only to a $3\times 3$ matrix, the third line describing the projection of the source plane into the observer plane. We establish that this generates an additional shear component. However since this effect is second order in the peculiar velocity of the source, it is not large enough to be observed. 

Secondly, in the usual derivation the shear and the convergence are expressed as functions of the source conformal time.
However conformal time is not an observable quantity. Hence, in this paper we calculate the shear and the convergence
as functions of the source redshift, which is observable.
We show that this modification generates new contributions to the convergence. 
Whereas most of those terms can be safely neglected with respect to the standard one, we
establish that the contribution of the source peculiar velocity is important, especially for redshifts $z\lesssim1$.
More particularly, for surveys in which the sources are situated at redshift $0.5$, we expect a modification of order $50\,\%$ relative
to the standard results. For larger redshifts of the source the effect of peculiar motion decreases, whereas the standard
term increases. However at redshift 1, we still expect the velocity term to be $\sim1\,\%$ of the standard term, \ie 
measurable by the SKA. Furthermore, we show that the transformation from conformal time to redshift does not affect the shear component.
As a consequence, the relation between the shear and the convergence is
modified by peculiar velocities.

The paper is organized as follows: in Sec.~\ref{sec:mag} we derive a general formula for the magnification 
matrix valid in (nearly) arbitrary geometries. In Sec.~\ref{sec:pert} we apply this formula to a perturbed Friedmann Universe.
In Sec.~\ref{sec:shear}, we investigate in detail the shear component of the magnification matrix.
Finally, in Sec.~\ref{sec:kappa} we calculate the convergence component and we determine its angular power spectrum. 
We investigate also the relation between cosmic shear and convergence.

{\it Notation:} We denote four-vectors with Greek indices, $k^\al$. Three-dimensional vectors are denoted bold face~$\bk$,
or with Latin indices~$k^i$. We use the metric signature $(-,+,+,+)$.

\section{Magnification matrix}
\label{sec:mag}

We consider an inhomogeneous and anisotropic Universe with geometry $ds^2=g_{\mu\nu}dx^\mu dx^\nu$. We are interested in 
the propagation of a light beam in this arbitrary spacetime.
We follow the derivation presented in~\cite{seitz}.
We denote by $\varphi$ the phase of the light beam.
The wave vector is then given by $k_\alpha=-\nabla_\alpha\varphi$. We construct the deviation vector field $\dx^\alpha$ 
connecting two neighboring rays. Since all the rays of the beam have the same phase, the deviation vector satisfies $\dx^\al k_\al=0$.
Furthermore it obeys the geodesic deviation equation \cite{SEF} 
\be
\label{eq:deviation}
\frac{D^2 \dx^\al (\la)}{D\la^2}=R^\al_{\beta\gamma\delta}k^\beta k^\gamma \dx^\delta~,
\ee
where $\la$ is an affine parameter along the geodesics, $\frac{D}{D\la}\equiv~k^\al\nabla_\al$ represents the 
covariant derivative along the geodesics,  and $R^\al_{\beta\gamma\delta}$ is the Riemann tensor associated with
the metric $ds^2$. 

\begin{figure}[ht]
\centerline{\epsfig{figure=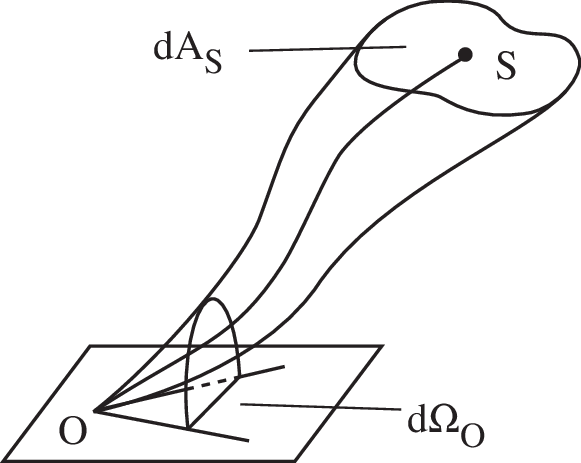,height=4cm}}
\caption{ \label{fig:beam} A light beam emitted by a galaxy at spacetime position $S$ and received by an observer at $O$.
At the observer position, the plane normal to the observer four-velocity is indicated.}
\end{figure}

We consider the case of a light beam emitted by a galaxy at spacetime position $S$ and received by an observer at $O$
(see Fig.~\ref{fig:beam}). 

We denote by $v_O$ the observer velocity and $v_S$ the source velocity. The photon energy measured at the source, respectively at
the observer is 
\bea
\om_S&=&-k_\al(\la_S)v^\al_{S}~,\\
\om_O&=&-k_\al(\la_O)v^\al_{O}~.
\eea
The solution of equation (\ref{eq:deviation}) is then given by \cite{sachs}
\be
\label{eq:jacobi}
\dx^\al(\la_S)=J^\al_{\,\,\beta}(\la_S)\dt^\beta(\la_O),
\ee
\be
\hbox{where} \hspace{0.2cm}\dt^\al(\la)\equiv\frac{1}{\om_O}k^\beta\nabla_\beta\dx^\al(\la)~.
\ee
Here $J^\al_{\,\,\beta}(\la_S)$ is the Jacobi map, that relates the deviation vector at the source
$\dx^\al(\la_S)\equiv \dx^\al_S$ to the angular vector
at the observer $\dt^\al(\la_O)\equiv~\dt^\al_O$. As we shall see, these two four-vectors belong
to two different two-dimensional planes, orthogonal to the source peculiar velocity (respectively the observer peculiar velocity)
and the photon direction at the source (respectively at the observer).

Each ray can be parameterized by its affine parameter $\la$ and three other parameters $y^i$ that label the ray. 
This parameterization is not unique and one can therefore make changes of the form
\be
\label{rep}
y^i=g^i(\tilde{y}^j)\,,\hspace{1cm} \la=\tilde{\la}+h(\tilde{y}^j)~.
\ee
Under this reparameterization the connection vector $\dx^\al$ transforms as~\cite{SEF}
\be
\delta\tilde{x}^\al=\dx^\al+k^\al\delta h~.
\ee

At the source, we can therefore choose a parameterization such that $\dx^\al_S v_{S\al}=0$. Moreover, $\dx^\al_S k_\al(\la_S)=0$
induces $\dx^\al_S n_{S\al}=0$, where $n_S$ is the photon
direction at the source: $n_S=\frac{1}{\om_S}k(\la_S)-v_S$. Hence $\dx^\al_S$ lives in a two-dimensional subspace orthogonal to 
the source velocity and to the photon direction. It lies consequently in the plane of the galaxy.
  
In addition to the choice of $g^i$ and $h$ in equation~(\ref{rep}), there is another degree of freedom corresponding to
the scales of the affine parameter on the different rays. We can
therefore require that at $O$, $v_{O \al}k^\al~=~v_{O\al}\frac{dx^\al}{d\la}=-\om_O$, for all rays.
With this choice, we get $\dx^\al\nabla_\al\om_O~=~0$, which gives $\dt_O^\al v_{O\al}~=~0$. Furthermore, we can easily show that
$\dt^\al(\la)k_\al(\la)=0$, which implies $\dt^\al_O n_{O\al}~=~0$, where $n_O$ is the photon
direction at the observer: $n_O~=~\frac{1}{\om_O}k(\la_O)~-~v_O$.
Hence $\dt^\al_O$ lives in a two-dimensional subspace orthogonal to 
the observer velocity and to the photon direction. It lies consequently in
the plane of the observer.

Therefore the Jacobi map $J^\al_{\,\,\beta}(\la_S)$ in equation~(\ref{eq:jacobi}) determines how the surface of the galaxy represented by
$\dx^\al_S$, is deformed into the observer angular vector $\dt^\al_O$, during propagation in an arbitrary spacetime.
It maps hence the surface of the galaxy to its
angular image at the observer.

We now want to relate this map to the magnification matrix defined in~\cite{seitz}.  At the observer we construct an orthonormal basis 
$\big(E_1^\al(\la_O),E_2^\al(\la_O),n_O^\al(\la_O),v_O^\al(\la_O)\big)$.
We parallel transport this basis along the geodesic. The subspace defined by $\big(E_1^\al(\la),~E_2^\al(\la)\big)$ is called the {\it screen}
adapted to $v_O^\al(\la)$ and $k^\al(\la)$.

We can write the deviation vector $\dx^\al_S$ and the angular vector $\dt^\al_O$ in this basis. Using the fact that $\dx^\al k_\al=0$,
we have
\bea
\dx^\al(\la)&=&-\xi_1(\la)E_1^\al(\la)-\xi_2(\la)E_2^\al(\la)\\
&&-\xi_0(\la)\big[n_O^\al(\la)+v_O^\al(\la)\big]\nonumber\\
&=&-\xi_1(\la)E_1^\al(\la)-\xi_2(\la)E_2^\al(\la)-\frac{\xi_0(\la)}{\om_O}k^\al(\la)~.\nonumber
\eea
Moreover, $\dt_O^\al k_\al(\la_O)=\dt_O^\al v_{O\al}=0$ implies
\be
\dt^\al_O=-\theta_1(\la_O)E_1^\al(\la_O)-\theta_2(\la_O)E_2^\al(\la_O)~.
\ee
In this basis equation (\ref{eq:jacobi}) becomes then
\be
\label{eq:jac3}
\left(\begin{array}{c}\xi_1\\ \xi_2\\ \xi_0 
\end{array}\right)(\la_S)=
\left(\begin{array}{ccc}&&\\ &\hat{J}^i_j(\la_O,\la_S)&\\ && 
\end{array}\right)\cdot
\left(\begin{array}{c}\theta_1\\ \theta_2\\ 0
\end{array}\right)(\la_O)~,
\ee
where
\bea
\lefteqn{\hat{J}^i_j(\la_O,\la_S)=}\\
&& \hspace{-0.4cm}\left(\begin{array}{ccc}E_{1\al}(\la_S)J^\al_\beta(\la_S)E_1^\beta(\la_O)&E_{1\al}(\la_S)J^\al_\beta(\la_S)E_2^\beta(\la_O)&0 \nonumber \\
E_{2\al}(\la_S)J^\al_\beta(\la_S)E_1^\beta(\la_O)&E_{2\al}(\la_S)J^\al_\beta(\la_S)E_2^\beta(\la_O)&0\nonumber\\
k_{\al}(\la_S)J^\al_\beta(\la_S)E_1^\beta(\la_O)&k_{\al}(\la_S)J^\al_\beta(\la_S)E_2^\beta(\la_O)&0
\end{array}\right)
\eea
$\hat{J}^i_{\, j}(\la_O,\la_S)$ is proportional to the magnification matrix of lens theory, i.e. the gradient of the
lens map.

Indeed, the lens map relates angles at the observer $\mbox{\boldmath $\theta$}_O$ to angles at the source $\bt_S$
\be
\bt_O\hspace{0.2cm}\rightarrow \hspace{0.2cm} \bt_S= \bt_O+\bal~.
\ee
The gradient of this map, $\mA^i_{\, j}$ is called the magnification matrix 
\be
\mathcal{A}^i_{\, j}=\delta^i_{\, j}+\frac{\dd \al^i}{\dd \theta^j_{O}}~.
\ee
For small angles, we can write $\bt_S=\mA\cdot\bt_O$, and consequently 
\be
\bxi_S=(\la_O-\la_S)\bt_S=(\la_O-\la_S)\mA\cdot\bt_O~.
\ee
This implies 
\be
\hat{J}(\la_O,\la_S)=(\la_O-\la_S)\mA~.
\ee
Usually, the component $\xi_0(\la_S)$ of the connection vector is neglected, and equation (\ref{eq:jac3}) becomes
\be
\left(\begin{array}{c}\xi_1\\ \xi_2 
\end{array}\right)(\la_S)= \Bigg(D^i_{\, j}(\la_O,\la_S)\Bigg)\cdot
\left(\begin{array}{c}\theta_1\\ \theta_2
\end{array}\right)(\la_O)~,
\ee
where $D^i_{\, j}$ is the $2\times 2$ submatrix of $\hat{J}^i_j$. The shear and the convergence are then extracted from $D^i_{\, j}$
(see e.g.~\cite{seitz}).

However, in general $D^i_{\, j}$ is not sufficient to relate $\bxi_S$ (or equivalently $\bt_S$) to $\bt_O$. 
The third line of the matrix $\hat{J}$ does indeed not vanish and therefore it plays a role in the deformation of the source plane.
This is due to the fact that even if the connection vector $\bxi_S$ at the source and the angular
vector $\bt_O$ at the observer are both two-dimensional, they do not live in parallel planes. The connection vector $\bxi_S$
belongs indeed to the plane normal to $v_S$ and $n_S$, whereas 
the angular vector $\bt_O$ belongs to the plane normal to $v_O$ and $n_O$. If $v_S\neq~v_O$,
these two planes are different.

Hence if we choose $\big(E_1^\al(\la_O),~E_2^\al(\la_O)\big)$ as a basis on the observer plane,
the parallel transported vectors  $\big(E_1^\al(\la_S),~E_2^\al(\la_S)\big)$ do {\it not} form a basis of the source plane.
The connection vector $\bxi_S$ will then have a component along $k^\al$, that we have to
take into account when calculating the shear and convergence of galaxies. The difference
between the two-dimensional plane of the source and the one of the observer comes mainly from
the difference between the peculiar velocity of the source and the observer. There is an additional effect, due to parallel transport,
which is however very small. Hence, if $v_S=v_O$, we can approximate the two planes as equal and we recover the
$2\times 2$ magnification matrix $D^i_{\, j}$.

In section~\ref{sec:pert}, we calculate explicitly the full matrix $\hat{J}$ in a perturbed Friedmann Universe (where
in general $v_S\neq~v_O$)  and
we show that the third line of the matrix $\hat{J}$ induces an additional shear effect, which is however negligible
since its amplitude is of second order in the source velocity.

\section{The magnification matrix in a perturbed Friedmann Universe}
\label{sec:pert}

We now calculate explicitly  the magnification matrix in a Friedmann Universe with scalar perturbations. In
longitudinal (or Newtonian) gauge the metric is given by
\be
g_{\mu\nu}dx^\mu dx^\nu = a^2\left[ -(1+2\Psi)d\eta^2
+(1-2\Phi)\ga_{ij}dx^idx^j\right] ~.
\ee
For perfect fluids like ordinary matter, dark matter and radiation, the metric perturbations $\Psi$ and $\Phi$ are
equal. Hence we assume in the sequel $\Phi=\Psi$. We restrict ourselves to a
spatially flat universe ($K=0$), so that $\ga_{ij}=\de_{ij}$.

Furthermore, since lightlike geodesics are not affected by conformal transformations,
we perform the calculation in the metric 
\begin{equation}
\tilde g_{\mu\nu}dx^\mu dx^\nu = \frac{1}{a^2} g_{\mu\nu}dx^\mu
dx^\nu ~.
\end{equation}
We can then easily relate the magnification matrix in the two metrics by remembering that 
angles are not affected by conformal transformations, but distances
scale with the conformal factor $a$. Hence 
\be
\hat{J}(\la_O,\la_S)= a_s \tilde{\hat J}(\la_O,\la_S)~.
\ee
We calculate therefore the magnification matrix related to the metric 
\be
\tilde{g}_{\mu\nu}dx^\mu dx^\nu =  -(1+2\Psi)d\eta^2
+(1-2\Psi)\delta_{ij}dx^idx^j ~,
\ee
and then we simply multiply the result by $a_S$ to obtain the correct magnification matrix in a perturbed Friedmann Universe.

\begin{figure}[ht]
\centerline{\epsfig{figure=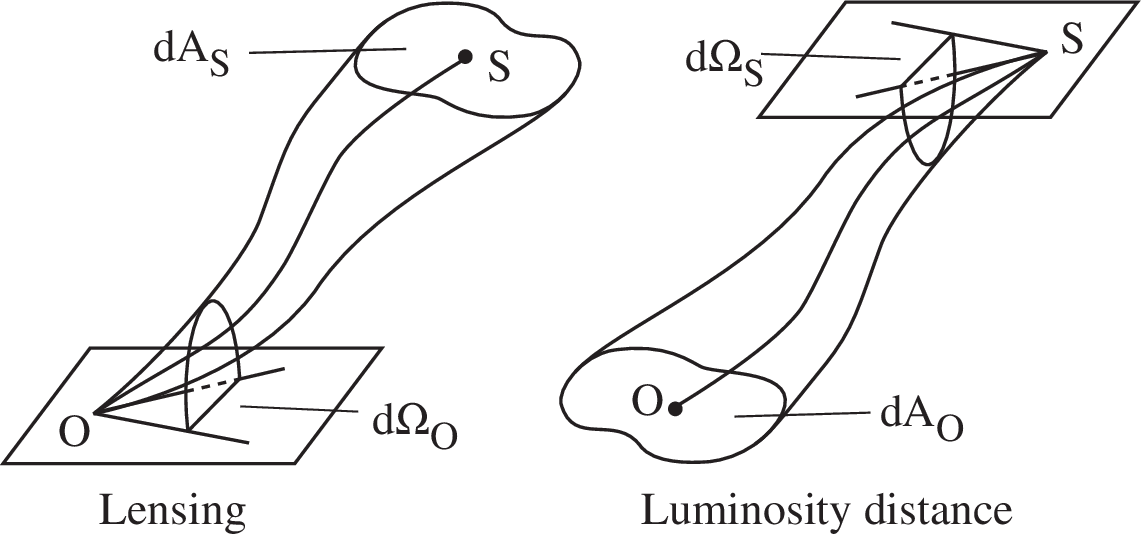,height=4cm}}
\caption{ \label{fig:lensing} On the left the "lensing" Jacobi map relates a surface at the source $dA_S$ to a solid angle at the observer $d\Om_O$, and on 
the right the "luminosity distance" Jacobi map relates a surface at the observer $dA_O$ to a solid angle at the source $d\Om_S$. The two maps are linked by equation
(\ref{eq:dL_lens}). }
\end{figure}

In \cite{luminosity}, we have calculated the Jacobi map associated with the luminosity distance in a perturbed Friedmann Universe, which
relates angles at the source $\dt_S^\al$ to distances at the observer $\dx_O^\al$
\be
\dx^\al_O=J^{(d_L)\,\al}_{\hspace{0.8cm}\beta}(\la_S,\la_O)\dt_S^\beta~.
\ee
This situation is the mirror of the lensing situation considered here (see equation~(\ref{eq:jacobi}) and Fig.~\ref{fig:lensing}).
The two maps can therefore easily be related~\cite{SEF} by
\be
\label{eq:dL_lens}
J^\al_{\,\,\beta}(\la_O,\la_S)=-\frac{\om_O}{\om_S}J^{(d_L)\hspace{0.1cm}\al}_{\hspace{0.5cm}\beta}(\la_S,\la_O)~.
\ee

Using $J^{(d_L)\,\al}_{\hspace{0.8cm}\beta}$ from~\cite{luminosity}, we find 
\be
\label{eq:mag}
\hat{J}^i_{\, j}(\eta_O,\eta_S)=-a_S(\eta_O-\eta_S)
\left(\begin{array}{ccc}1-\kappa-\gamma_1&-\gamma_2&0\\
-\gamma_2&1-\kappa+\gamma_1&0\\
\bw\cdot\bE_1&\bw\cdot\bE_2&0
\end{array}\right)
\ee
with
\bea
\label{eq:ga1}\gamma_1&=&\!\!\!\!\int_{\eta_S}^{\eta_O}\!\!\!\!\!\!d\eta\frac{(\eta-\eta_S)(\eta_O-\eta)}{\eta_O-\eta_S}\pa_i\pa_j\Psi (E_1^iE_1^j\!-\!E_2^iE_2^j)\\
\label{eq:ga2}\gamma_2&=&2\int_{\eta_S}^{\eta_O}d\eta \frac{(\eta-\eta_S)(\eta_O-\eta)}{\eta_O-\eta_S}\pa_i\pa_j\Psi E_1^iE_2^j \\
\kappa&=&\int_{\eta_S}^{\eta_O}\!\!\!\!d\eta \frac{(\eta-\eta_S)(\eta_O-\eta)}{\eta_O-\eta_S}\left(\nabla^2-n^in^j\partial_i\partial_j\right)\Psi \nonumber\\
&&-\frac{4}{(\eta_O-\eta_S)}\int_{\eta_S}^{\eta_O}\!\!\!d\eta\, \Psi-2\int_{\eta_S}^{\eta_O}\!\!\!d\eta \frac{\eta-\eta_S}{\eta_O-\eta_S}\,\dot{\Psi}\nonumber\\
&&+\bv_O\cdot\bn+\Psi_S+2\Psi_O\phantom{\Big(}\\
\label{eq:w}\bw&=&\bv_S-\bv_O-\int_{\eta_S}^{\eta_O}\!\!d\eta\,\nabla\Psi~.
\eea
Note that a similar derivation of $\gamma_1, \gamma_2$ and $\kappa$ in the context of CMB lensing is presented in~\cite{lewis}.

We have written here the magnification matrix as a function of conformal time $\eta$.
However, $\eta$ is not an observable quantity. What we do measure is the redshift
of the galaxy, which is also affected by perturbations, $z_S=\bar{z}_S~+~\delta z_S$. 
Now
\be
\hJ(\eta_S)=\hJ(\eta(\bar z_S))\equiv \hJ(\bar z_S) =  \hJ(z_S) - \frac{d}{d\bar z_S}
 \hJ(z_S)\de z_S ~.
\ee
Furthermore,
\bea
  \frac{d}{d\bar z_S} \hJ( z_S) &=& \frac{1}{1+z_S}\left(\frac{1}{\apt_S(\eta_O-\eta_S)}-1 \right)\hJ
+  \mbox{1st order }
   \nonumber \\
\eea
and~(see \cite{luminosity})
\bea
\label{eq:dz}
\de z_S &=& (1+z_S)\Big[\Psi_S - \Psi_O
+2\int_{\eta_S}^{\eta_O}\!\!d\eta\,\bn\cd\bnabla\Psi\nonumber\\
& +& (\bv_O-\bv_S)\cd\bn\Big]~. 
\eea
Hence $\kappa$ becomes
\bea
\label{eq:kappa}
\kappa&=&\int_{\eta_S}^{\eta_O}\!\!\!\!d\eta \frac{(\eta-\eta_S)(\eta_O-\eta)}{\eta_O-\eta_S}\nabla_\perp^2\Psi-\frac{2}{(\eta_O-\eta_S)}\int_{\eta_S}^{\eta_O}\!\!\!\!d\eta\, \Psi\nonumber\\
&+& 2\left(1-\frac{1}{\apt_S(\eta_O-\eta_S)}\right)\int_{\eta_S}^{\eta_O}\!\!\!d\eta\, \dot\Psi \nonumber\\
&+&\left(1-\frac{1}{\apt_S(\eta_O-\eta_S)}\right) (\bv_S-\bv_O)\cd\bn +\bv_O\cd\bn \nonumber\\
&+&\left(1-\frac{1}{\apt_S(\eta_O-\eta_S)}\right) (\Psi_S-\Psi_O)+\Psi_S~,
\eea
where $\nabla_\perp^2\Psi \equiv \big(\nabla^2-n^in^j\pa_i\pa_j+2(\eta_O-\eta)^{-1}n^i\pa_i\big)\Psi $ is the transverse Laplacian. The terms $\gamma_1, \ga_2$ and $\bw$ are not affected at linear order.

The complete relativistic magnification matrix in a perturbed Friedmann Universe defined in
equations~(\ref{eq:mag})-(\ref{eq:ga2}),(\ref{eq:w}) and (\ref{eq:kappa})
contains therefore additional terms to the standard one. 
In the next section, we investigate the effect of the third line of the matrix on the shape of a galaxy,
and more particularly on the shear. 
And in section~\ref{sec:kappa}, we discuss in more detail the different contributions to the
convergence $\kappa$. We determine that the only term which can be relevant is the one involving the galaxy peculiar velocity.

\section{The shear}
\label{sec:shear}

In this section, we study the effect of the third line of the magnification matrix on the shape of a galaxy.
In order to simplify the calculation, we restrict ourselves to the peculiar velocity contribution.
This means that we consider a homogeneous and isotropic Friedmann Universe, but we allow for nonzero peculiar velocity
of the source~$\bv_S$ and of the observer~$\bv_O$.
The magnification matrix becomes then
\bea
\label{eq:jacv}
\lefteqn{\hat{J}^i_{\, j}(\eta_O,\eta_S)=}\\
&&-\frac{\eta_O-\eta_S}{1+z_S}
\left(\begin{array}{ccc}1-\kappa&0&0\\
0&1-\kappa&0\\
(\bv_S-\bv_O)\bE_1\hspace{0.2cm}&(\bv_S-\bv_O)\bE_2\hspace{0.2cm}&0
\end{array}\right)~,\nonumber
\eea
where
\be
\kappa=\left(1-\frac{1}{\apt_S(\eta_O-\eta_S)}\right)(\bv_S-\bv_O)\cd\bn+\bv_O\cd \bn~.
\ee
The usual shear components $\gamma_1$ and $\ga_2$ vanish, but as we will see the third line generates an additional shear deformation. 

From equation~(\ref{eq:jac3}) and (\ref{eq:jacv}), we find 
\be
\label{eq:xi}
\xi_0=\frac{v_1\xi_1}{1-\kappa}+\frac{v_2\xi_2}{1-\kappa}~,
\ee
where $v_1\equiv(\bv_S-\bv_O)\cdot \bE_1$ and $v_2\equiv(\bv_S-\bv_O)\cdot\bE_2$. At first order in $\bv_S$ and $\bv_O$, equation~(\ref{eq:xi}) becomes
\be
\label{eq:plane}
\xi_0-v_1\xi_1-v_2\xi_2=0~,
\ee
that represents the equation of a two-dimensional plane. This reflects directly the fact that the galaxy, described by $\bxi_S$
does not belong to the parallel transported observer plane: $\big(E_1^\al(\la_S),~E_2^\al(\la_S)\big)$,
but it rather belongs to the plane defined by equation~(\ref{eq:plane}).

We now assume that the galaxy is a disc of radius $r$. Hence in spherical coordinate
\be
\xi_1=r\sin\theta\cos\phi~,\hspace{0.09cm} \xi_2=r\sin\theta\sin\phi~,\hspace{0.09cm}\xi_0=r\cos\theta~.
\ee
Combined with equation~(\ref{eq:plane}), this gives 
\bea
\xi_1(\phi)&=&\pm r\frac{t(\phi)}{\sqrt{1+t^2(\phi)}}\cos\phi~,\nonumber\\
\xi_2(\phi)&=&\pm r\frac{t(\phi)}{\sqrt{1+t^2(\phi)}}\sin\phi~,
\eea
\be
\hbox{with}\hspace{0.3cm}t(\phi)=\frac{1}{v_1\cos\phi+v_2\sin\phi}~.
\ee
The $+$~sign is for $\phi\in[0,\pi/2] \cup [\pi,3\pi/2]$ and the $-$~sign for $\phi\in[\pi/2,\pi] \cup [3\pi/2,2\pi]$. 

The observer measures
\bea
\label{eq:theta}
\theta_1(\phi)&=&\frac{\xi_1(\phi)}{1-\kappa}=\pm \frac{r}{1-\kappa}\frac{t(\phi)}{\sqrt{1+t^2(\phi)}}\cos\phi~,\nonumber\\
\theta_2(\phi)&=&\frac{\xi_2(\phi)}{1-\kappa}=\pm \frac{r}{1-\kappa}\frac{t(\phi)}{\sqrt{1+t^2(\phi)}}\sin\phi~,
\eea
where we have already removed the monopole contribution $\frac{\eta_O-\eta_S}{1+z_S}$.

We now show that $\big(\theta_1(\phi),\theta_2(\phi)\big)$ describes an ellipse, and we determine its semiaxis $a$ and $b$.
We consider, without loss of generality, the case where $v_2=0$. This simply means that we align $\bE_1$ on $\bv_S-\bv_O$. 
We then determine $a$ and $b$ such that 
\be
\frac{\theta_1^2(\phi)}{a^2}+\frac{\theta_2^2(\phi)}{b^2}=1~, \hspace{0.5cm} \forall\, \phi~.
\ee
Using equation~(\ref{eq:theta}) for $\theta_i(\phi)$, we find 
\be
b=\frac{r}{1-\kappa}\hspace{0.3cm}\hbox{and}\hspace{0.3cm}a=\frac{r}{1-\kappa}\left(1+v_1^2\right)^{-1/2}~.
\ee

Hence we see that the third line of the magnification matrix $\hJ$ deforms a disc into an ellipse and consequently
it induces an additional shear effect. Since this new contribution does not derive from a scalar potential, as the usual component, it can generate B-modes.
Peculiar motion constitutes therefore an intrinsic source of B-modes in cosmic shear. Moreover, contrary to the B-modes' contribution from source redshift clustering~\cite{bmodes} that peaks at small scales, the velocity contribution is expected to peak at rather large scales, where peculiar velocity correlations are larger.
However,
this effect is second order in the velocity difference $v_1$. It is therefore too small to be detected by current and future experiments and consequently to be responsible for the observed B-modes \cite{bobs}.
In the following we can therefore safely neglect it with respect to the standard shear components $\ga_1$
and $\ga_2$ and reduce the magnification matrix to the usually considered $2\times 2$ submatrix. 

Note that in this calculation we did not
take into account the effect of the potential, which induces an integrated Sachs-Wolf term, $\int_{\eta_S}^{\eta_O}\!\! d\eta\nabla\Psi$ 
in equation~(\ref{eq:w}). However, this term is much smaller than the peculiar velocity contribution (see e.g.~\cite{luminosity}) and can be neglected. 
Hence, the effect of the third line on the shape of the galaxy reduces to a  purely kinematic effect,
which can also be understood as length contraction of the galaxy in the velocity direction. 

\section{Convergence}
\label{sec:kappa}

\subsection{The velocity contribution to the convergence.}

We evaluate now the convergence $\kappa$. The various terms in equation~(\ref{eq:kappa}) are
very similar to the one affecting the luminosity distance perturbations. In~\cite{luminosity}, we have estimated all these terms
and we found two dominant contributions: the lensing term and the source peculiar velocity term.
We neglect hence the other terms here. Among them, one finds the effect of peculiar velocity of the lenses, which corresponds to the gradient of the potential in equations~(\ref{eq:w}) and (\ref{eq:dz}). Those terms which may be relevant in intermediate lensing (see e.g.~\cite{serono} and references therein) are completely subdominant in weak lensing.
Therefore in the following we restrict our calculations to the two components
\be
\kappa=\kappa_\Psi+\kappa_\bv~,
\ee 
where
\bea
\label{eq:kp} \kappa_\Psi&=&\int_{\eta_S}^{\eta_O}\!\!\!\!d\eta \frac{(\eta-\eta_S)(\eta_O-\eta)}{\eta_O-\eta_S}\nabla_\perp^2\Psi~,\\ 
\label{eq:kv} \kappa_\bv&=&\left(1-\frac{1}{\apt_S(\eta_O-\eta_S)}\right) (\bv_S-\bv_O)\cd\bn+\bv_O\cd \bn~.\nonumber\\
\eea
The convergence can be measured through the modifications it induces on the galaxy number density at a given flux.

Let us introduce the magnification 
\be
\mu=\frac{1}{\det\mA}\simeq  1+2\kappa, \hspace{0.4cm}\mbox{if}\hspace{0.2cm} \kappa, \gamma_1, \gamma_2 \ll 1.
\ee  
The magnification modifies the size of an observed source: $d\Om_O=\mu d\Om_S $, where $d\Om_S$ is the true angular size
of the source and 
$d\Om_O$ is the solid angle measured by the observer, \ie the size of the image. The lensing term $\kp$ is always positive 
and consequently it always magnifies the source. 
On the contrary, the velocity term $\kv$ can be either positive or negative and it therefore either magnifies or demagnifies
the source. The sign of $\kv$ depends on the sign of $\bv_S\cdot\bn$ and on the sign of 
$g(z_S)\equiv\left(1-\frac{1}{\apt_S(\eta_O-\eta_S)}\right)$.

To fix the ideas, let us consider a galaxy moving toward us. Then 
$\bv_S\cd\bn>0$, since $\bn$ represents the photon direction, which points to the observer. 
Moreover, the sign of $g(z_S)$ depends on the redshift of the source $z_S$. 
In a $\La$CDM Universe with $\Om_m=0.24$ and $\Om_\La=0.76$, we have $g(z)<0$  for $z<1.7$ and $g(z)>0$ for $z>1.7$. 
Hence for small $z$ the surface is demagnified ($\mu_\bv=1+2\kv<1$) and for large $z$ it is magnified ($\mu_\bv>1$). 
This is caused by the change in redshift induced by the source peculiar velocity.
Indeed, for a fixed redshift,
a source moving toward us is more distant (in conformal time for example) than a source with null peculiar velocity.
This generates two opposite effects on the observed solid angle. On one hand, a distant galaxy is observed under a smaller solid angle.
And on the other hand, since its conformal time is smaller, its scale factor is also smaller. Consequently, the image
experiences more expansion when coming to us. At small redshift the first effect dominates, leading to a demagnification
of the source, whereas at large redshift the second effect dominates and the source is magnified. At $z = 1.7$,
both effects compensate, leaving the size of the source unchanged.
The situation is then simply reversed for a source moving away from us. 

We evaluate now the effect of magnification (or demagnification) on the galaxy number density.
We consider $\bar{n}(f)df$ unlensed galaxies per unit solid angle at a redshift $z_S$ and with a flux in the range $[f,f+df]$.
The magnification modifies the flux measured by the observer, since it modifies the observed galaxy surface.
It modifies also the solid angle of observation and hence the number of galaxy 
per unit of solid angle. These two effects combine to give a galaxy number overdensity~\cite{sdss}
\be
\label{eq:overdens}
\de^\mu_g=\frac{n(f)-\bar{n}(f)}{\bar{n}(f)}\simeq1+2\big(\al-1\big)(\kp+\kv)~.
\ee
Here $\al\equiv -N'(>f_c)f_c/N(f_c)$, where $N(>f_c)$ is the number of galaxies brighter than $f_c$ and $f_c$ is
the flux limit adopted. Hence $\al$ is an observable quantity~\cite{sdss,pen}.

Recent measurements of the galaxy number overdensity $\de^\mu_g$ are reported in~\cite{sdss}. The challenge in those measurements is to 
eliminate intrinsic clustering of galaxies, which induces an overdensity $\de_g^{cl}$ much larger than $\de_g^\mu$. One
possibility to separate these two effects is to correlate
galaxy number overdensities at widely separated redshifts. One can then measure $\langle \de_g^\mu(z_S)\de_g^{cl}(z_\Sp)\rangle$,
where $z_S$ is the redshift
of the sources and $z_\Sp<z_S$ is the redshift of the lenses. 
Recently, the authors of~\cite{pen} proposed to remove close pairs of galaxies
at the same redshift from the signal in order to eliminate $\de_g^{cl}$ . This allows then to measure $\langle \de_g^\mu(z_S)\de_g^{\mu}(z_\Sp)\rangle$, either for $z_S\neq z_\Sp$, or for
$z_S=z_\Sp$, \ie for galaxies situated at the same redshift but in different directions. This method requires of course to know
precisely the redshift of the galaxies.

In the following we study in detail
the contribution of peculiar motion to $\langle\de_g^\mu(z_S)\de_g^{\mu}(z_\Sp)\rangle$.
The two components of the convergence $\kp$ and $\kv$ (and consequently the galaxy number overdensity)
are functions of redshift $z_S$ and direction of observation $\bn$. We can therefore determine the angular power spectrum
\be
\langle \de^\mu_g(z_S,\bn)\de^\mu_g(z_\Sp,\bn')\rangle=\sum_{\ell}\frac{2\ell+1}{4\pi}C^\mu_\ell(z_S,z_\Sp)P_\ell(\bn\cdot \bn')~.
\ee
The coefficients $C^\mu_\ell(z_S,z_\Sp)$ contain two kinds of terms induced by $\langle \kp \kp\rangle$ and $\langle \kv \kv\rangle$.
The cross-term
$\langle \kv \kp\rangle$  vanishes since $\kp$ contains only Fourier modes with a wave vector $\mathbf{k}_\perp$ perpendicular to the line of sight (see eq.~(\ref{eq:kp})),
whereas $\kv$ selects modes with wave vector along the line of sight (eq.~(\ref{eq:kv})). 

The velocity contribution is
\bea
\label{eq:deltav}
\lefteqn{\langle \de_g^\bv(z_S,\bn)\de_g^\bv(z_\Sp,\bn')\rangle=}\\
&&4(\al_S-1)(\al_\Sp-1)\langle(\bv_S\cd\bn)(\bv_\Sp\cd\bn')\rangle\cdot\nonumber\\
&&\!\left(1-\frac{1}{\apt_S(\eta_O-\eta_S)}\right)
\cdot\left(1-\frac{1}{\apt_\Sp(\eta_O-\eta_\Sp)}\right)~.\nonumber
\eea
Here we neglect the peculiar velocity of the observer $\bv_O$ which gives rise only to a dipole.

We use the Fourier transform convention 

\bea
\bv(\bx,\eta)&=&\frac{1}{(2\pi)^3}\int d^3k\, \bv(\bk,\eta)e^{i\bk\bx}~,\\
\bv(\bk,\eta)&=&\int d^3x\, \bv(\bx,\eta)e^{-i\bk\bx}~.
\eea
With the solution of the continuity equation \cite{Dod}
\be
\label{eq:velocity}
\bv(\bk,\eta)=i\frac{D'(a)}{D(a)}\frac{\bk}{k^2}\delta(\bk,a)~,
\ee
where $\delta(\bk,\eta)$ is the density contrast, $D(a)$ is the growth function, and
$D'(a)$ its derivative with respect to $\eta$, we find 
\bea
\lefteqn{C^{\bv}_\ell(z_S,z_\Sp)=\frac{16\pi\delta_H^2(\al_S-1)(\al_\Sp-1)}{(H_0\eta_O)^2D^2(a=1)}
\frac{D'(z_S)D'(z_\Sp)}{H_0^2}\cdot}\nonumber\\
&&\left(1-\frac{1}{\apt_S(\eta_O-\eta_S)}\right)\left(1-\frac{1}{\apt_\Sp(\eta_O-\eta_\Sp)}\right)\cdot\nonumber\\
&&\eta_O^2\int dk k T^2(k)j_\ell'(k(\eta_O-\eta_S))j_\ell'(k(\eta_O-\eta_\Sp))
\eea
Here $\delta_H$ is the density contrast at horizon and $T(k)$ is the transfer function defined through~\cite{Dod}
\be
\Psi(\bk,a)=\frac{9}{10}\Psi_p(\bk)T(k)\frac{D(a)}{a}~,
\ee
where $\Psi_p(\bk)$ is the primordial power spectrum.

We want to compare this contribution with the usual contribution coming from the potential $\kp$
\bea
\lefteqn{C^{\Psi}_\ell(z_S,z_\Sp)=\frac{36\pi\delta_H^2(\al_S-1)(\al_\Sp-1)\Om_m^2}{D^2(a=1)}\cdot}\\
&&D(z_S)D(z_\Sp)(1+z_S)(1+z_\Sp)\cdot\int dk k^3 T^2(k)\nonumber\\
&&\int_{\eta_S}^{\eta_O}d\eta W(\eta,\eta_S)\Big( j_\ell(k(\eta_O-\eta))+j''_\ell(k(\eta_O-\eta))\Big)\cdot\nonumber\\
&&\int_{\eta_\Sp}^{\eta_O}d\eta'W(\eta',\eta_\Sp)\Big(j_\ell(k(\eta_O-\eta'))+j''_\ell(k(\eta_O-\eta'))\Big)~,\nonumber
\eea
where $W(\eta,\eta_S)=\frac{(\eta-\eta_S)(\eta_O-\eta)}{(\eta_O-\eta_S)}$ is the lensing kernel.
We evaluate $\cv$ and $\cp$ in a $\La$CDM Universe with $\Om_m~=~0.24$, $\Om_\La~=~0.76$ and $n_s~=~1$. 
$\delta_H$ can be measured from the Sachs-Wolf plateau~\cite{ruth}. We find $\delta_H=6\cdot 10^{-5}$~\cite{WMAP5}. 
We use the BBKS transfer function~\cite{BBKS} (see also~\cite{eisenstein} which proposes a similar fit). 

\begin{figure}[t]
\centerline{\epsfig{figure=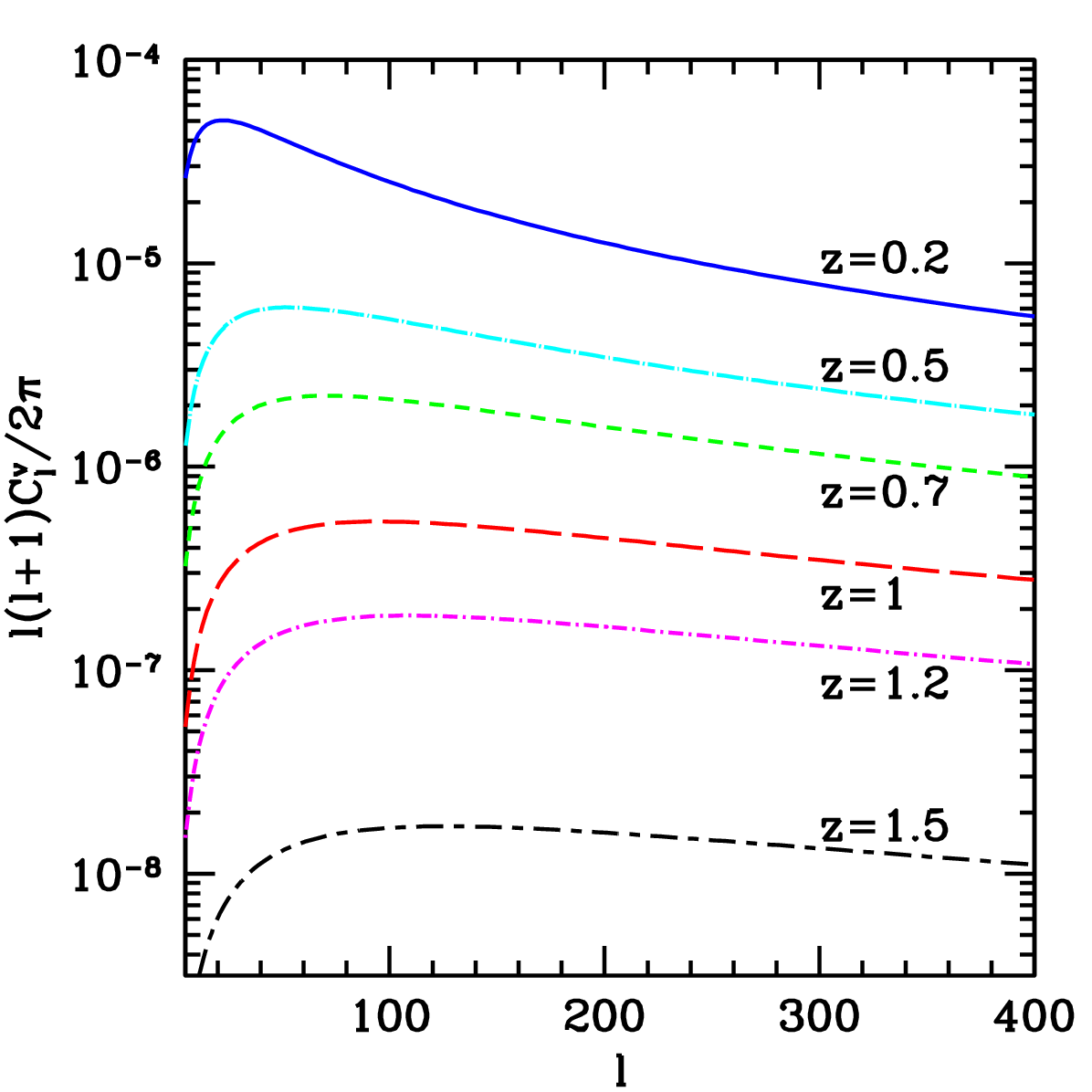,height=7.7cm}}
\caption{ \label{fig:clv} The velocity contribution $\cv$ as a function of $\ell$, for redshift (from top to bottom)
$z=0.2, 0.5, 0.7, 1, 1.2$ and $1.5$. We see that $\cv$ decreases with redshift in this redshift range.  }
\end{figure}

\begin{figure}[t]
\centerline{\epsfig{figure=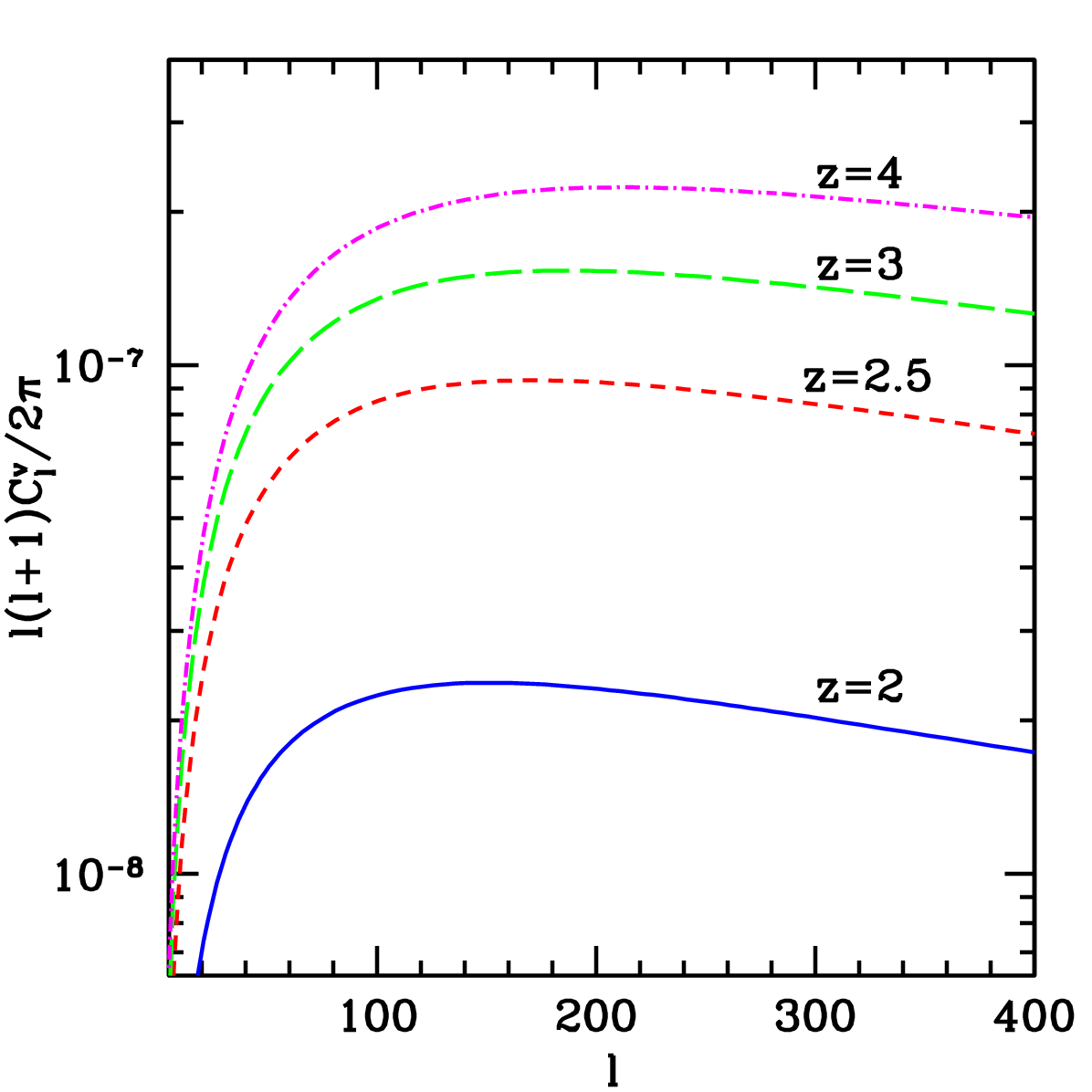,height=7.7cm}}
\caption{ \label{fig:clv2} The velocity contribution $\cv$ as a function of $\ell$, for redshift (from top to bottom)
$z=4,3,2.5$ and $2$. We see that $\cv$ increases with redshift in this redshift range.}
\end{figure}

\begin{figure}[t]
\centerline{\epsfig{figure=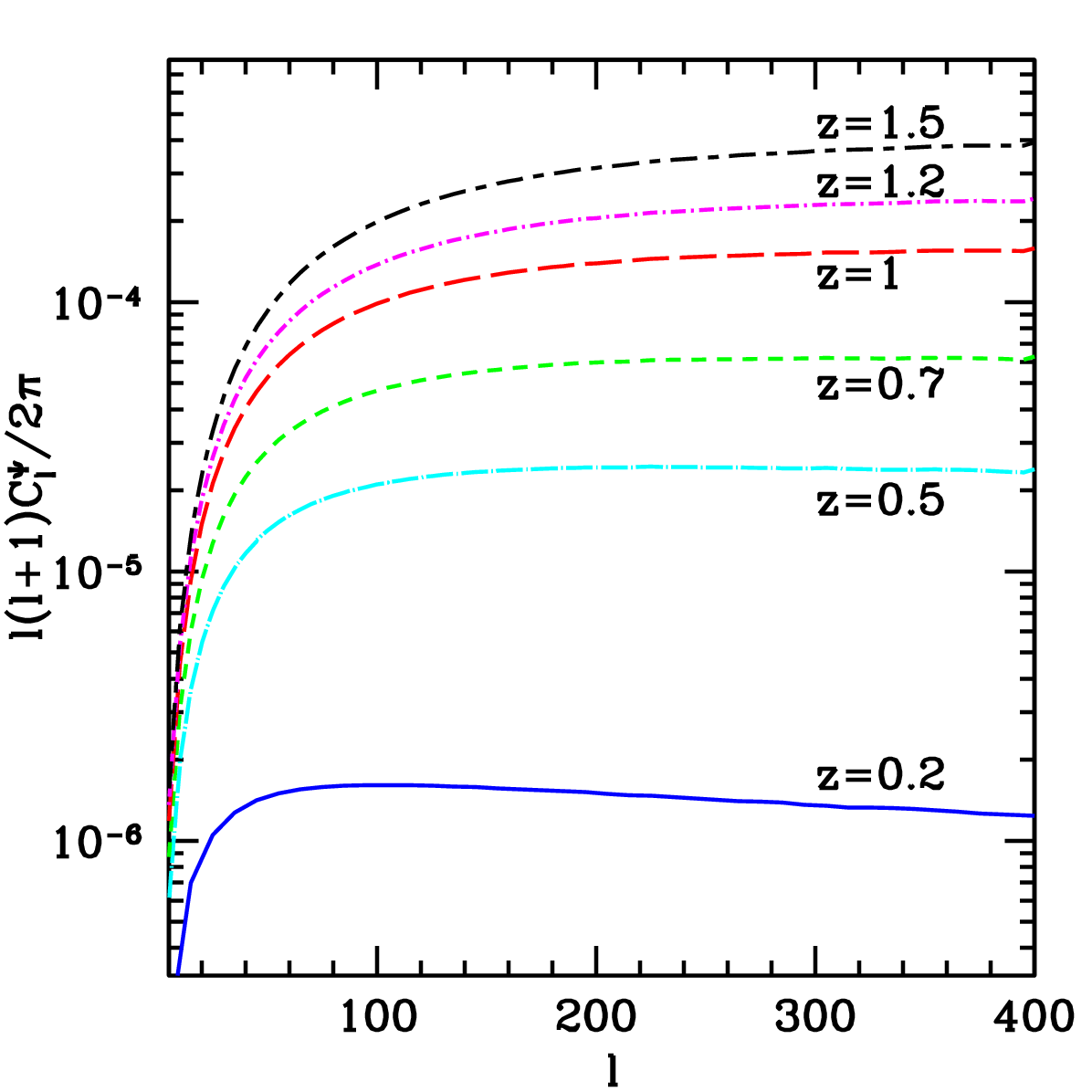,height=7.7cm}}
\caption{ \label{fig:lens} The potential contribution $\cp$ as a function of $\ell$, for redshift (from top to bottom)
$z=1.5, 1.2, 1, 0.7, 0.5 $ and $0.2$.}
\end{figure}

\begin{figure}[t]
\centerline{\epsfig{figure=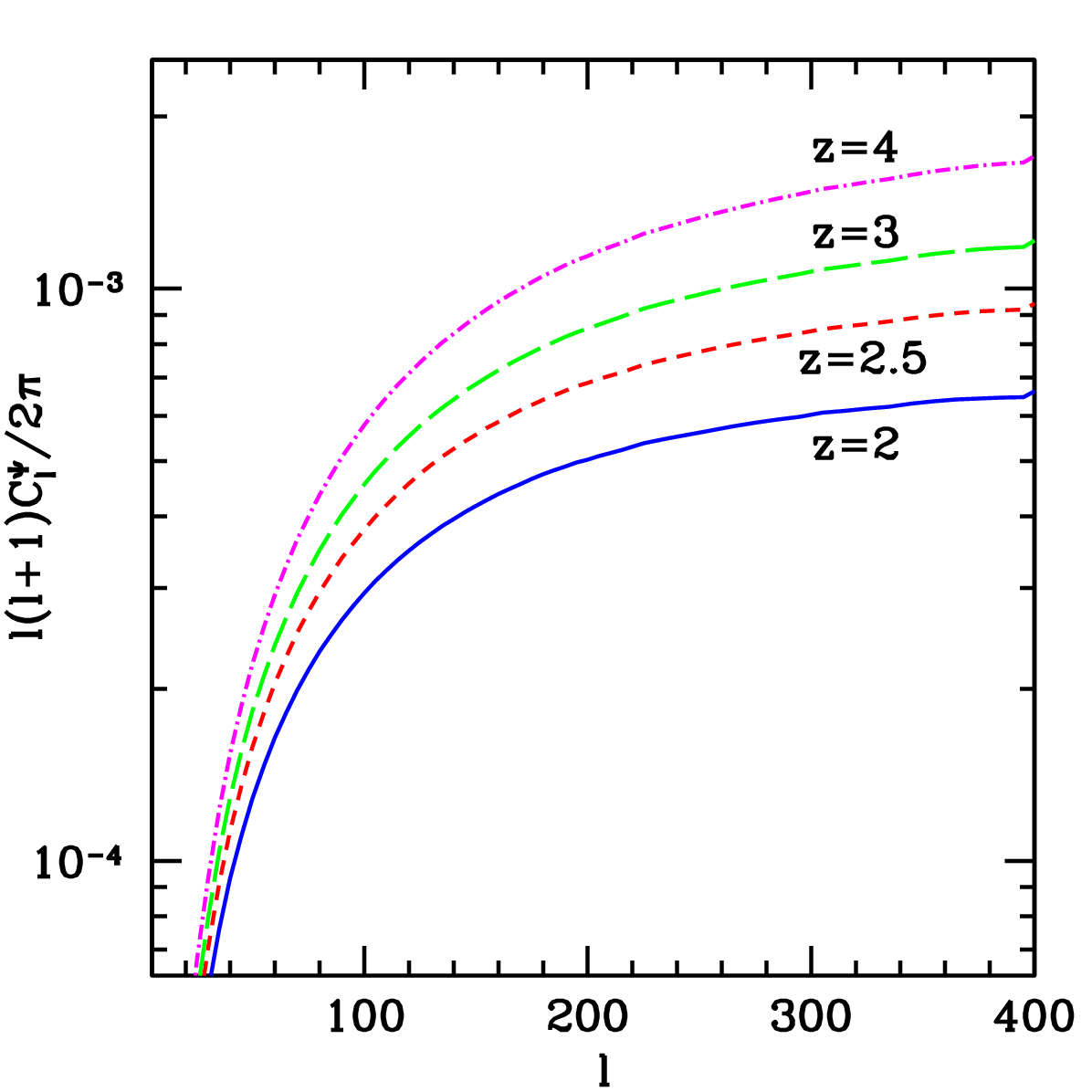,height=7.7cm}}
\caption{ \label{fig:lens2} The potential contribution $\cp$ as a function of $\ell$, for redshift (from top to bottom)
$z=4,3,2.5$ and $2$.
}
\end{figure}

In Fig.~\ref{fig:clv}-\ref{fig:lens2}, we plot $C_\ell^\bv$ and $C_\ell^\Psi$ for different redshifts, but with $z_S=z_\Sp$. 
The amplitude of $\cv$ and $\cp$ depends on $(\alpha-1)^2$, which varies with the redshift of the source, the
flux threshold adopted, and the sky coverage of the experiment~\cite{pen}. Since this term influences
$\cv$ and $\cp$ in the same way  we do not 
include it in our plot. Generally, at small redshifts, $(\al-1)$ is smaller than 1 and consequently the amplitude
of both $\cv$ and $\cp$ is slightly reduced, whereas at large redshifts $(\al-1)$ tends to be larger than 1 and to amplify
$\cv$ and $\cp$~\cite{pen}. However, the general features of the curves and more importantly the ratio between $\cv$ and $\cp$
are not affected by $(\al-1)$.

Figures~\ref{fig:clv} and \ref{fig:clv2} show that $\cv$ peaks at rather small $\ell$, between $30$ and $150$ depending on the redshift.
The behavior of $\cp$ is different, since it increases with $\ell$ (Fig.~\ref{fig:lens} and \ref{fig:lens2}). 
Hence, peculiar motion of galaxies generates additional correlations that peak at relatively large angle
$\theta\sim 70-360 $ arcmin. It is therefore important to have large sky surveys to detect this effect.

\begin{figure}[ht]
\centerline{\epsfig{figure=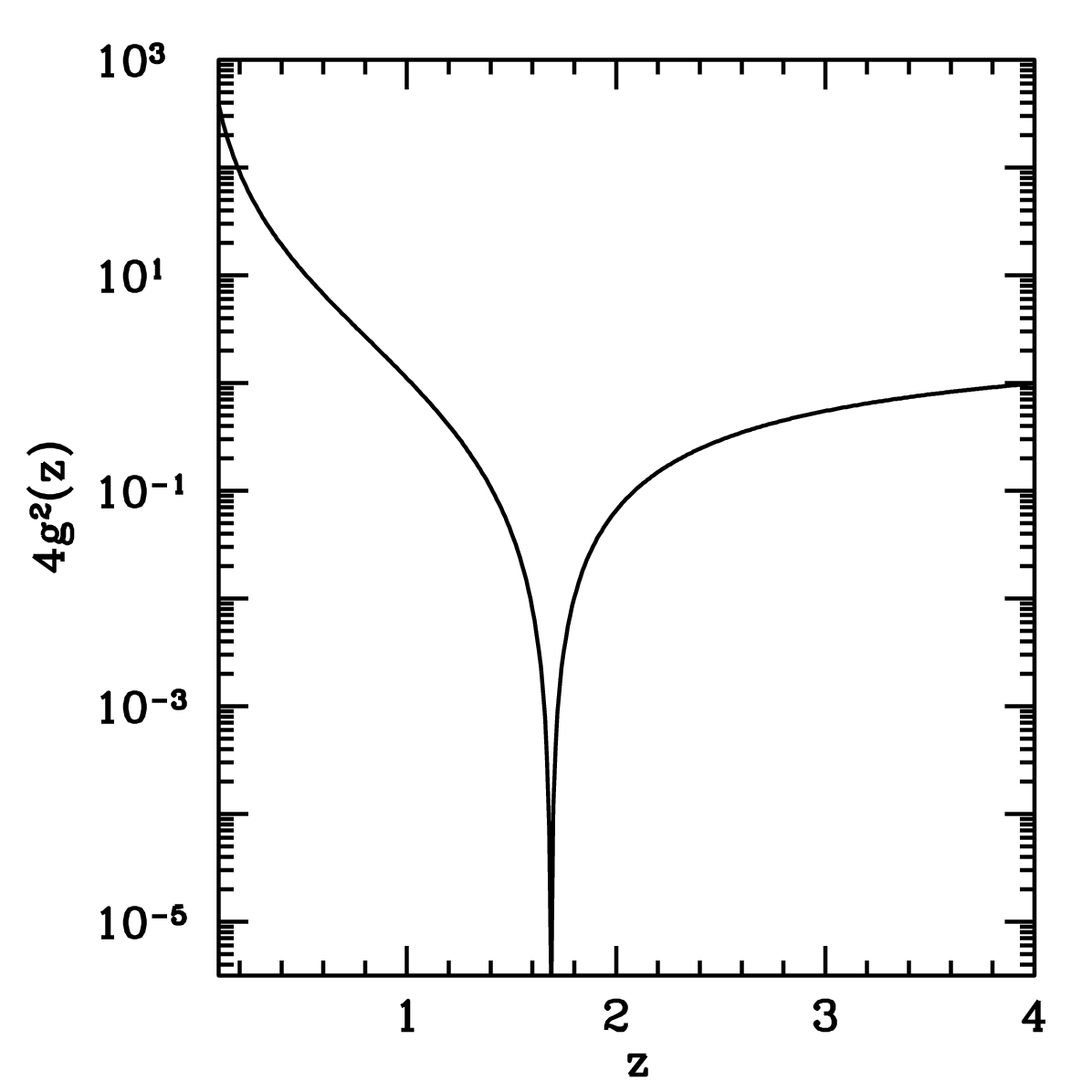,height=7.7cm}}
\caption{ \label{fig:g2} The prefactor $4g^2(z)$ of the velocity term $\cv$ as a function of redshift.}
\end{figure}

The relative importance of $\cv$ and $\cp$ de\-pends stron\-gly on the redshift of the source. 
At small redshift, $z_S=0.2$, the velocity contribution is about $5\cdot 10^{-5}$ and is hence larger than the lensing contribution which reaches $10^{-6}$. 
At redshift $z_S=0.5$, $\cv$ is about 50\,$\%$ of $\cp$, whereas at redshift $z_S=1$, it is about 1\,$\%$ of $\cp$. 
Then at redshift $z_S=1.5$ and above, $\cv$ becomes very small with respect to $\cp$: $\cv\lesssim10^{-4}\,\cp$.
This fast decrease of $\cv$ is a consequence of the behavior of $g^2(z)$ plotted in Fig.~\ref{fig:g2}. We see
that at redshift $z_S=1.7$, $\cv$ vanishes, due to the fact that $\kv=0$, as explained above. Finally,
for $z_S>1.7$, $g(z)$ and consequently $\cv$ start to increase as
a function of $z_S$ (Fig.~\ref{fig:clv2}). However, the lensing 
term $\cp$ increases faster than the velocity term and hence the ratio $\cv/\cp$ stays small, even at redshift
4 where it reaches a few $10^{-4}$. 

We have also calculated $\cv(z_S,z_\Sp)$ for $z_S\neq z_\Sp$. We found that it decreases very rapidly with the redshift difference,
and hence it gives a negligible contribution to correlations between different redshift bins. On the contrary,
$\cp(z_S,z_\Sp)$ is not very sensitive to the redshift difference and induces therefore correlations at
different redshifts.

To summarize, we see from Fig.~\ref{fig:clv} and \ref{fig:clv2} that a lensing survey must satisfy three criteria
in order to detect the velocity contribution $\cv$. First it has to cover a rather large part of the sky,
since the velocity contribution peaks at angles between 
$70$ arcmin and $6$ degrees  (depending on the redshift of the source). Ongoing and future surveys fulfill
this requirement. For example, the CFHTLS Wide survey has already spanned 57 square degrees of the sky (over three independent fields)
and allowed consequently to measure the two-point shear statistics from 1 arcmin to 4 degrees.
And in the near future, measurements will be extended up to 8 degrees~\cite{Fu}. Future surveys are even more ambitious.
For example, Euclid and the SKA plan to cover up
to $20'000$ square degrees. The requirement of a wide field survey is also of great importance for shear measurements.
It is necessary in order to, on one hand, beat cosmic variance, and on the other hand probe accurately the linear regime
of perturbations.

Second, in order to measure precisely the galaxy number overdensity, one needs to remove intrinsic clustering.
One way to perform this, consists simply of removing from the signal close pairs of galaxies at the same redshift, as proposed
in~\cite{pen}. This requires one to measure the redshift of the galaxies.
The SKA offers
here a great advantage, since it can measure with high precision 
the redshift of galaxies from 21cm emission line wavelength.
Most other future lensing surveys will deliver photometric redshift measurements of the observed galaxies.
This is indeed also crucial for cosmic shear observation. First it allows, as for magnification measurements, to remove close
pairs of galaxies, in order to reduce the systematic noise due to intrinsic alignment of galaxies in the same 
cluster~\cite{ellipticity}. Second, redshift information permits one to perform 3D cosmic shear analysis.

Finally, Fig.~\ref{fig:clv} shows that the velocity contribution is mainly important at redshift $z\lesssim1$.
Hence an ideal survey should cover redshifts from 0.2 to 1. This is well within the range of ongoing and future experiments.

The accuracy with which the velocity contribution can actually be measured as well as the exact redshifts range where it is detectable
depend of course of the precision of the
survey. In~\cite{pen}, measurements of the magnification autocorrelations by the 
SKA are discussed in detail. 
The autocorrelations are calculated for redshifts $z\gtrsim 1$. Since at redshift~$1$, our velocity contribution is of order~$1\,\%$ of the standard term,
it should be measurable by the SKA if, as expected, an accuracy of~$1\,\%$ is reached. At larger redshifts,
the velocity contribution decreases quickly and will hence not be detectable. However, the SKA should
be able to measure the velocity contribution at smaller redshifts. Indeed, at small redshift $z\simeq 0.2$,
the velocity contribution is rather large, of the order $5\cdot10^{-5}$. Hence between redshift $0.2\leq z\leq 1$,
the velocity contribution
should be measurable. Between redshift $0.5\leq z\leq 1.5$, a large number of galaxies ($\gtrsim10^4 \, \hbox{deg}^{-2}$) are expected
to be detected by the SKA, hence good statistic should be achieved in this redshift range.  Of course, the amplitude of the signal
is weighted by $(\al-1)^2$, which is not included in Fig.~\ref{fig:clv} and \ref{fig:clv2}. At very small redshift, $\al-1$ can be small,
and even negative. If $(\al-1)\sim -0.5$, the amplitude
is for example reduced by 4. The value of $\al$ depends on the features of the survey: sky coverage, selection threshold,
redshift of the galaxies, etc... Hence one can optimize this value to detect the velocity contribution.

\subsection{Nonlinear contribution}

Until here, we have only considered peculiar velocities from linear theory, where equation~(\ref{eq:velocity}) holds. 
However, on small scales, galaxies' peculiar velocities do not obey linear theory anymore. Typically, galaxies in a cluster
can have peculiar velocities of the order $v_S\sim~1000$~km/s~\cite{peebles}. Equation~(\ref{eq:deltav}) becomes then
\be
\langle \de_g^\bv(z_S,\bn)\de_g^\bv(z_S,\bn')\rangle\simeq(\al_S-1)^2 4g^2(z_S)
\cdot10^{-5}~.
\ee

The amplitude depends on the prefactor $4g^2(z)$ plotted in Fig.~\ref{fig:g2},  and on  $(\alpha-1)^2$. 
Hence the nonlinear contribution is large up to redshift $z\simeq 1.4$, where it is still of the order of $10^{-6}$,
if we take $\alpha-1\simeq 1$. Then between $z=1.4$ and $z=2.1$ it becomes
very small. But for $z\gtrsim 2.1$ the signal is again non-negligible and reaches even $10^{-5}$ at
$z=4$. However at those redshifts, we do not expect to find clusters of galaxies.

Moreover, the nonlinear velocity contribution is large only on very small scales and it should become negligible when one removes  close pairs of galaxies
at the same redshift from the signal, as proposed in~\cite{pen}.

\subsection{The reduced shear} 

In section \ref{sec:shear} we have established that peculiar motion contributes at second order to the shear and is therefore too small to be detected. However, since what is actually observed is not the shear, but the reduced shear \cite{reduced}
\be
g=\frac{\gamma}{1-\kappa}\simeq \gamma-\kappa\gamma \hspace{0.5cm}\mbox{for} \hspace{0.2cm} \kappa, \gamma \ll 1~,
\ee
it is necessary to investigate the effect of peculiar motion on $g$.

The reduced shear correlations are given by
\bea
\lefteqn{\langle g(z_S,\bn) g(z_{S'},\bn')\rangle=\langle \gamma(z_S,\bn) \gamma(z_{S'},\bn')\rangle}\\
&&-\langle\gamma(z_S,\bn) \gamma(z_{S'},\bn') \kappa(z_{S'},\bn')\rangle\nonumber\\
&&- \langle\gamma(z_S,\bn) \kappa(z_{S},\bn) \gamma(z_{S'},\bn')\rangle \hspace{0.2cm}+\hspace{0.2cm}\mbox{higher order}~.\nonumber
\eea
The corrections $\langle\gamma(z_S,\bn) \gamma(z_{S'},\bn') \kp(z_{S'},\bn')\rangle$ have been computed in~\cite{white}. They are large enough to impact on 
cosmological parameter estimation from future experiments. 
However, one can show that the velocity induced corrections $\langle\gamma(z_S,\bn) \gamma(z_{S'},\bn')\kv(z_{S'},\bn')\rangle$ vanish.
Indeed $\gamma(z_S,\bn)$ contains only Fourier modes with a wave vector $\mathbf{k}_\perp$ perpendicular to the line of sight (see eqs.~(\ref{eq:ga1}) and (\ref{eq:ga2})). On the other hand $\kv(z_S,\bn)$ selects modes with wave vector along the line of sight (eq.~(\ref{eq:kv})). Hence correlations between $\kv$ and $\gamma$ vanish and 
we have 
\bea
\lefteqn{\langle\gamma(z_S,\bn) \gamma(z_{S'},\bn')\kv(z_{S'},\bn')\rangle=}\\
&&\langle\gamma(z_S,\bn) \gamma(z_{S'},\bn')\rangle\langle\kv(z_{S'},\bn')\rangle=0~.
\eea
Therefore peculiar motion does not affect the reduced shear correlations at first order in $\kv$.

\subsection{Relation between shear and convergence}
\label{sec:comp}

In the usual description of weak lensing, the shear and the convergence satisfy a functional relation. 
They are indeed both related to second order derivatives of the potential along the line of sight.
In the flat sky approximation,  we have 
\bea
\kappa_{\Psi}(\eta_S,\bb)&=& \Delta_{\beta}\hat\Psi(\eta_S,\bb)~, \label{eq:k_flat}\\
\gamma_1(\eta_S,\bb)&=& (\nabla_{\beta_1}\nabla_{\beta_1} -\nabla_{\beta_2}\nabla_{\beta_2})\hat\Psi(\eta_S,\bb)\label{eq:g1_flat}~,\\
\gamma_2(\eta_S,\bb)&=&2\nabla_{\beta_1}\nabla_{\beta_2} \hat\Psi(\eta_S,\bb)~, \label{eq:g2_flat}
\eea
where 
\be
 \hat\Psi(\eta_S,\bb)=\int_{\eta_S}^{\eta_O}\!\!d\eta W(\eta,\eta_S)\Psi(\eta,\bb)~.
\ee
Here  $\bb=(\beta_1,\beta_2)$ is a two-dimensional angular vector describing the position of the source in the flat sky approximation, and $\nabla_{\beta}$
is the associated two-dimensional gradient. Equation~(\ref{eq:k_flat}) can be solved to express $\hat\Psi$ as a function of $\kappa_\Psi$. Inserting the result into equations~(\ref{eq:g1_flat}) and (\ref{eq:g2_flat}) leads to the following relation between $\kp$ and $\gamma$  (see e.g.~\cite{straumann})
\be
\gamma(\bb)=\gamma_1(\bb)+i\gamma_2(\bb)=\frac{1}{\pi}\int d^2\beta ' D(\bb-\bb')\kappa_\Psi(\bb')~,
\ee
where
\be
D(\bb)=\frac{\beta_2^2-\beta_1^2-2i\beta_1\beta_2}{|\bb|^4}~.
\ee
The two-dimensional Fourier transform of the shear and the convergence are therefore related through $\hat\gamma(\mathbf{q})=e^{2i\varphi}\hat\kp(\mathbf{q})$, where $\varphi$ is the polar angle of $\mathbf{q}$. Consequently, the two-dimensional power spectrum of the convergence, $P_{\kappa_\Psi}$ and of the shear, $P_\ga$ are equal.

A similar relation holds in the all sky calculation, where the angular power spectra satisfy~\cite{allsky}
\be
C_\ell^{\kappa_\Psi}=\frac{\ell(\ell+1)}{(\ell+2)(\ell-1)}C_\ell^\gamma~.
\ee

These relations are modified by the velocity contribution. Galaxies' peculiar velocities generate indeed a new contribution to the convergence, $\kv$ that is not related to the potential $\hat\Psi$.
On the other hand, it does not change the shear at first order. Hence, in the flat sky approximation the measured shear $\gamma$ and convergence $\kappa_{\rm obs}=\kp+\kv$ obey 
\be
\gamma(\bb)=\frac{1}{\pi}\int d^2\beta ' D(\bb-\bb')\big(\kappa_{\rm obs}(\bb')-\kv(\bb')\big)~. 
\ee
Since the correlations between $\kv$ and $\kp$ vanish, the power spectra satisfy
\be
\label{eq:pgamma}
P_\gamma=P_{\kappa_{\rm obs}}-P_{\kv}~.
\ee

Similarly in the all sky calculation, we have
\be
\label{eq:cgamma}
\frac{\ell(\ell+1)}{(\ell+2)(\ell-1)}C_\ell^\gamma=C_\ell^{\kappa_{\rm obs}}-C_\ell^{\kv}=\frac{C_\ell^\mu-C_\ell^\bv}{4(\al-1)^2}~,
\ee
where we have used equation~(\ref{eq:overdens}) for the second equality sign.

Consequently, if one measures both the shear $C_\ell^\gamma$ and the magnification $C_\ell^\mu$ as functions of the redshift, equations~(\ref{eq:cgamma})
allows one to extract the peculiar velocity contribution $C_\ell^\bv$.
This provides a new way to measure directly peculiar velocities of galaxies.
Future surveys, like Euclid and the SKA should be able to deliver measurements of both the shear and the magnification
in the redshift range $0.2\leq z\leq 1$. Since at those redshifts the difference between the velocity contribution
and the potential contribution is at least of the percent level, it seems feasible to extract peculiar velocities from the measurements
of $C_\ell^{\mu}$ and $C_\ell^\gamma$.

A direct measure of the peculiar motion of galaxies would be extremely interesting, since it allows one to determine the underlying
matter power spectrum. Moreover, since the velocity contribution $\cv$ peaks at rather large angles, it provides a measurement
of the velocity field well inside the linear regime, where the relation between peculiar velocity and matter overdensity is
simply given by equation~(\ref{eq:velocity}).
A more careful analysis is of course necessary to determine if this method can compete with actual or future observations
of peculiar velocities, from galaxy surveys (see e.g.~\cite{velocity} and references therein), supernova surveys~\cite{SN}  or from
the kinetic Sunyaev-Zel'dovich effect in the Cosmic Microwave Background (see e.g.~\cite{SZ,pablo} and references therein).

\section{Conclusions}

In this work we have studied the effect of peculiar velocity on weak gravitational lensing.
We have derived a general formula for the shear and the convergence, that takes into account all
relativistic effects of linear perturbation theory. We have identified a new important contribution 
generated by the peculiar motion of galaxies.

We have shown that the shear component is affected only at second order by peculiar velocity.
Consequently this contribution does not affect  cosmic shear in a measurable way.
However, we have found that the effect of peculiar velocity on the convergence is important and cannot be neglected,
especially for redshift $z\lesssim1$. At small redshifts $z\lesssim0.4$, the peculiar velocity contribution
is even larger than the lensing contribution.
Hence measurements of the convergence (or more particularly of the magnification) at small and intermediate redshifts are
affected by the peculiar motion of galaxies. In the redshift range $0.2\leq z\leq 1$ we expect an effect large enough to be detected by
future galaxy surveys, like, for example, the SKA and Euclid.

We have also shown that peculiar motion modifies the relation between cosmic shear and convergence. One way to measure
the effect of peculiar velocity would hence consist in measuring both the shear and the convergence at the same redshifts, and
in comparing them to extract the velocity contribution. This provides hence a new way of measuring the velocity field directly.

Finally, we have found that the redshift dependence of the velocity term is mainly given by the prefactor $g^2(z)$, plotted in 
Fig.~\ref{fig:g2}. Since this function depends directly on cosmological parameters, it would be very interesting
to find a way to extract it from the data, in order to constrain cosmological models.

\acknowledgments
I would like to thank Marc-Olivier Bettler, Ruth Durrer, Pedro Ferreira, and Antony Lewis for useful and stimulating
discussions and comments on the first draft of this manuscript. I also thank Nick Kaiser and the authors of~\cite{roy} for pointing out a mistake in the expression of the convergence eq.~(\ref{eq:kappa}). This work was supported by the Swiss National Science foundation.

\twocolumngrid

\end{document}